\documentclass[preprint,nofootinbib]{revtex4}
\usepackage[latin9]{inputenc}
\usepackage{textcomp}
\usepackage{amsmath}
\usepackage{amssymb}
\usepackage{esint}

\makeatletter
\@ifundefined{textcolor}{}
{%
 \definecolor{BLACK}{gray}{0}
 \definecolor{WHITE}{gray}{1}
 \definecolor{RED}{rgb}{1,0,0}
 \definecolor{GREEN}{rgb}{0,1,0}
 \definecolor{BLUE}{rgb}{0,0,1}
 \definecolor{CYAN}{cmyk}{1,0,0,0}
 \definecolor{MAGENTA}{cmyk}{0,1,0,0}
 \definecolor{YELLOW}{cmyk}{0,0,1,0}
}

\usepackage{amsfonts}
\usepackage{graphicx}
\providecommand{\U}[1]{\protect\rule{.1in}{.1in}}

\makeatother

\begin{document}

\title{Extension of the Poincaré group with half-integer spin generators:\\
 hypergravity and beyond}

\author{Oscar Fuentealba$^{1,2}$, Javier Matulich$^{1}$, Ricardo Troncoso$^{1}$}

\email{fuentealba, matulich, troncoso@cecs.cl}

\affiliation{$^{1}$Centro de Estudios Científicos (CECs), Casilla 1469, Valdivia,
Chile}

\affiliation{$^{2}$Departamento de F\'{i}sica, Universidad de Concepción, Casilla
160-C, Concepción, Chile.}

\preprint{CECS-PHY-15/02}
\begin{abstract}
An extension of the Poincaré group with
half-integer spin generators is explicitly constructed. We start discussing
the case of three spacetime dimensions, and as an application, it
is shown that hypergravity can be formulated so as to incorporate
this structure as its local gauge symmetry. Since the algebra admits
a nontrivial Casimir operator, the theory can be described in terms
of gauge fields associated to the extension of the Poincaré group
with a Chern-Simons action. The algebra is also shown to admit an
infinite-dimensional non-linear extension, that in the case of fermionic
spin-$3/2$ generators, corresponds to a subset of a contraction of
two copies of \textrm{WB}$_2$. Finally, we show how the Poincaré
group can be extended with half-integer spin generators for $d\geq3$
dimensions.
\end{abstract}
\maketitle

\section{Introduction}

Nowadays, we have the good fortune of witnessing the era in which
the simplest minimal realistic versions of supersymmetric field theories
are about to be either tested or falsified by the LHC. The underlying
geometric structure of these kind of theories, as well as most of
their widely studied extensions, relies on the super-Poincaré group
(see, e.g., \cite{FF-SUSY}, \cite{Sohnius-SUSY}, \cite{VN-SUGRA}).
According to the Haag-\L opusza\'{n}ski-Sohnius theorem \cite{HLS},
this is a consistent extension of the Poincaré group that includes
fermionic generators of spin $1/2$. Indeed, in flat spacetimes of
dimension greater than three, the addition of fermionic generators
of spin $s\geq3/2$ would imply that the irreducible representations
necessarily contained higher spin fields, which are known to suffer
from inconsistencies (see, e.g., \cite{Weinberg}, \cite{AD-ConsistencyProb},
\cite{BVHVN}, \cite{AD2}, \cite{Weinberg-Witten}, \cite{Porrati},
\cite{Bekaert-Boulanger-Sundell}). However, in three spacetime dimensions,
higher spin fields do not possess local propagating degrees of freedom,
and as a consequence, it is possible to describe them consistently
\cite{Blencowe}, \cite{BBS}, \cite{Vasiliev-HS}, \cite{Henneaux-Rey},
\cite{CFPT-HS3D}, \cite{HPTT-CPHS3D}, \cite{BHPTT-GBH3D} even on
locally flat spacetimes \cite{ABFGR-HSF3D}, \cite{GMPT-HSF3D}, \cite{GGRR-CPHSF3D},
\cite{MPTT-CPHSF3D}. Hence, in the latter context, since no-go theorems
about massless higher spin fields can be circumvented, it is natural
to look for an extension of the Poincaré group with fermionic half-integer
spin generators. Results along these lines have already been explored
in \cite{Hietarinta}. In what follows, we begin with the construction
of the searched for extension of the Poincaré group in the case of
spin $3/2$ generators, that for short, hereafter we dub it the ``hyper-Poincaré''
group. It is shown that the algebra admits a nontrivial Casimir operator
and, as an application, we explain how the hypergravity theory of
Aragone and Deser \cite{AD-3D-HYGRA} can be formulated so as to incorporate
the hyper-Poincaré group as its local gauge symmetry. Concretely,
we show how hypergravity can be described in terms of hyper-Poincaré-valued
gauge fields with a Chern-Simons action. The results are then extended
to the case of fermionic generators of spin $n+\frac{1}{2}$, as well
as to the minimal coupling of General Relativity with gauge fields
of spin $n+\frac{3}{2}$, so that super-Poincaré group and supergravity
are recovered for $n=0$. The hyper-Poincaré algebra is also shown
to admit an infinite-dimensional nonlinear extension that contains
the BMS$_{3}$ algebra, which in the case of spin-$3/2$ generators,
reduces to a subset of a suitable contraction of two copies of \textrm{WB}$_2$.
We conclude explaining how the hyper-Poincaré group is extended to
the case of $d\geq3$ dimensions.

\section{Fermionic spin-$3/2$ generators}

In three spacetime dimensions, the nonvanishing commutators of the
Poincaré algebra can be written as
\begin{equation}
\left[J_{a},J_{b}\right]=\varepsilon_{abc}J^{c}\quad,\quad\left[J_{a},P_{b}\right]=\varepsilon_{abc}P^{c}\,.
\end{equation}
The additional fermionic generators are assumed to transform in an
irreducible spin-$3/2$ representation of the Lorentz group, so that
they are described by ``$\Gamma$-traceless'' vector-spinors that
fulfill $Q^{a}\Gamma_{a}=0$, where $\Gamma_{a}$ stand for the Dirac
matrices. Their corresponding commutation rules with the Lorentz generators
are then given by 
\begin{equation}
\left[J_{a},Q_{\alpha b}\right]=\frac{1}{2}\left(\Gamma_{a}\right)_{\,\,\,\,\alpha}^{\beta}Q_{\beta b}+\varepsilon_{abc}Q_{\alpha}^{c}\,.\label{eq:JQ}
\end{equation}
Therefore, requiring consistency of the closure as well as the Jacobi
identity, implies that the only remaining nonvanishing (anti-) commutators
of the algebra read 
\begin{equation}
\left\{ Q_{\alpha}^{a},Q_{\beta}^{b}\right\} =-\frac{2}{3}\left(C\Gamma^{c}\right)_{\alpha\beta}P_{c}\eta^{ab}+\frac{5}{6}\varepsilon^{abc}C_{\alpha\beta}P_{c}+\frac{1}{6}(C\Gamma^{(a})_{\alpha\beta}P^{b)}\,,\label{eq:QQ}
\end{equation}
where $C$ is the charge conjugation matrix %
\footnote{In our conventions, the Minkowski metric $\eta_{ab}$ is assumed to
follow the ``mostly plus'' convention, and the Levi-Civita symbol
fulfills $\varepsilon_{012}=1$. Round brackets stand for symmetrization
of the enclosed indices, without the normalization factor, e. g.,
$X^{\left(a\right|}Y^{b}Z^{\left|c\right)}=X^{a}Y^{b}Z^{c}+X^{c}Y^{b}Z^{a}$.
It is also useful to keep in mind the Fierz expansion of the product
of three Dirac matrices, given by $\Gamma^{a}\Gamma^{b}\Gamma^{c}=\varepsilon^{abc}+\eta^{ab}\Gamma^{c}+\eta^{bc}\Gamma^{a}-\eta^{ac}\Gamma^{b}$.
Afterwards, the presence of the imaginary unit \textquotedblleft $i$\textquotedblright{}
in the product of real Grassmann variables is because we assume that
$\left(\theta_{1}\theta_{2}\right)^{*}=-\theta_{1}\theta_{2}$.%
}. It is then simple to verify that apart from $I_{1}=P^{a}P_{a}$,
the algebra admits another Casimir operator given by 
\begin{equation}
I_{2}=2J^{a}P_{a}+Q_{\alpha}^{a}C^{\alpha\beta}Q_{\beta a}\,,\label{eq:Casimir}
\end{equation}
which implies the existence of an invariant (anti-) symmetric bilinear
form, whose only nonvanishing components are of the form
\begin{eqnarray}
\left\langle J_{a},P_{b}\right\rangle  & = & \eta_{ab}\,,\nonumber \\
\left\langle Q_{\alpha}^{a},Q_{\beta}^{b}\right\rangle  & = & \frac{2}{3}C_{\alpha\beta}\eta^{ab}-\frac{1}{3}\varepsilon^{abc}(C\Gamma_{c})_{\alpha\beta}\,.\label{eq:Brackets}
\end{eqnarray}
It is worth highlighting that the inclusion of the higher spin generators
$Q_{\alpha}^{a}$ does not jeopardize the causal structure, since
there is no need to enlarge the Lorentz group.

\subsection{Hypergravity}

In order to describe a massless spin-$\frac{5}{2}$ field minimally
coupled to General Relativity, let us consider a connection 1-form
that takes values in the hyper-Poincaré algebra described above, which
reads 
\begin{equation}
A=e^{a}P_{a}+\omega^{a}J_{a}+\psi_{a}^{\alpha}Q_{\alpha}^{a}\,,\label{eq:Connection}
\end{equation}
where $e^{a}$, $\omega^{a}$ and $\psi_{a}^{\alpha}$ stand for the
dreibein, the dualized spin connection ($\omega^{a}=\frac{1}{2}\varepsilon^{abc}\omega_{bc}$),
and the $\Gamma$-traceless spin-$\frac{5}{2}$ field ($\Gamma^{a}\psi_{a}=0$),
respectively. The components of the field strength $F=dA+A^{2}$ are
then given by
\begin{equation}
F=R^{a}J_{a}+\tilde{T}^{a}P_{a}+D\psi_{a}^{\alpha}Q_{\alpha}^{a}\,,
\end{equation}
where the covariant derivative of the spin-$\frac{5}{2}$ field can
be written as $D\psi^{a}=d\psi^{a}+\frac{3}{2}\omega^{b}\Gamma_{b}\psi^{a}-\omega_{b}\Gamma^{a}\psi^{b}$,
and $R^{a}=d\omega^{a}+\frac{1}{2}\epsilon^{abc}\omega_{b}\omega_{c}$
is the dualized curvature 2-form. The hypercovariant torsion 2-form
then reads 
\begin{equation}
\tilde{T}^{a}:=T^{a}-\frac{3}{4}i\bar{\psi}_{b}\Gamma^{a}\psi^{b}\,,
\end{equation}
with $T^{a}=de^{a}+\varepsilon^{abc}\omega_{b}e_{c}$, and $\bar{\psi}_{a\alpha}=\psi_{a}^{\beta}C_{\beta\alpha}$
is the Majorana conjugate.

Note that under an infinitesimal gauge transformation $\delta A=d\lambda+\left[A,\lambda\right]$,
spanned by a hyper-Poincaré-valued zero-form given by $\lambda=\lambda^{a}P_{a}+\sigma^{a}J_{a}+\epsilon_{a}^{\alpha}Q_{\alpha}^{a}$,
the components of the gauge field transform according to 
\begin{eqnarray}
\delta e^{a} & = & D\lambda^{a}-\varepsilon^{abc}\sigma_{b}e_{c}+\frac{3}{2}i\bar{\epsilon}_{b}\Gamma^{a}\psi^{b},\nonumber \\
\delta\omega^{a} & = & D\sigma^{a},\label{eq:Gauge-trans}\\
\delta\psi^{a} & = & -\frac{3}{2}\sigma^{b}\Gamma_{b}\psi^{a}+\sigma_{b}\Gamma^{a}\psi^{b}+D\epsilon^{a}.\nonumber 
\end{eqnarray}

The invariant bilinear form \eqref{eq:Brackets} then allows to construct
a Chern-Simons action for the gauge field \eqref{eq:Connection},
given by
\begin{equation}
I=\frac{k}{4\pi}\int\left\langle AdA+\frac{2}{3}A^{3}\right\rangle ,\label{eq:CS-action}
\end{equation}
which up to a boundary term, reduces to 
\begin{equation}
I=\frac{k}{4\pi}\int2R^{a}e_{a}+i\bar{\psi}_{a}D\psi^{a}\,.\label{eq:AD-action}
\end{equation}
It is worth pointing out that, despite the action \eqref{eq:AD-action}
is formally the same as the one considered by Aragone and Deser in
\cite{AD-3D-HYGRA}, it does possess a different local structure.
Indeed, note that under local hypersymmetry transformations spanned
by $\lambda=\epsilon_{a}^{\alpha}Q_{\alpha}^{a}$, the nonvanishing
transformation rule for the spin connection considered in \cite{AD-3D-HYGRA},
agrees with ours only on-shell. Actually, by construction, as in the
case of supergravity \cite{BTZ-HDCS}, here the algebra of the local
gauge symmetries \eqref{eq:Gauge-trans} closes off-shell according
to the hyper-Poincaré group, without the need of auxiliary fields. 

In the case of negative cosmological constant, it can be seen that
hypergravity requires the presence of additional spin-4 fields \cite{CLW},
\cite{ZInoviev}, \cite{Hyper-AdS}.

\section{Fermionic generators of spin $\displaystyle{s=n+\frac{1}{2}}$}

In this case, the fermionic generators correspond to tensor-spinors
$Q_{\alpha}^{a_{1}\dots a_{n}}$, transforming in an irreducible representation
of the Lorentz group, so that they are completely symmetric in the
vector indices, as well as $\Gamma$-traceless, i. e., $Q^{a_{1}\dots a_{n}}\Gamma_{a_{1}}=0$.
These conditions imply that their anticommutation rules acquire a
somehow cumbersome expression, and it is then more convenient to write
the hyper-Poincaré algebra in the Maurer-Cartan formalism. The Maurer-Cartan
1-form is given by 
\begin{equation}
\Omega=\rho^{a}P_{a}+\tau^{a}J_{a}+\chi_{a_{1}\dots a_{n}}^{\alpha}Q_{\alpha}^{a_{1}\dots a_{n}},\label{eq:MC-form}
\end{equation}
where $\chi_{a_{1}\dots a_{n}}$ is $\Gamma$-traceless and completely
symmetric in the vector indices, which can be seen as a flat connection
that fulfills 
\begin{align}
d\tau^{a} & =-\frac{1}{2}\epsilon^{abc}\tau_{b}\tau_{c}\,,\label{eq:MC-algebra}\\
d\rho^{a} & =-\epsilon^{abc}\tau_{b}\rho_{c}+\frac{1}{2}\left(n+\frac{1}{2}\right)i\bar{\chi}_{a_{1}\dots a_{n}}\Gamma^{a}\chi^{a_{1}\dots a_{n}}\,,\nonumber \\
d\chi^{a_{1}\dots a_{n}} & =-\left(n+\frac{1}{2}\right)\tau_{b}\Gamma^{b}\chi^{a_{1}\dots a_{n}}+\tau_{b}\Gamma^{(a_{1}}\chi^{a_{2}\dots a_{n})b}\,.\nonumber 
\end{align}
Note that the Jacobi identity now translates into the consistency
of the nilpotence of the exterior derivative ($d^{2}=0$), which for
the algebra \eqref{eq:MC-algebra} is clearly satisfied.

The nontrivial Casimir operator now reads
\begin{equation}
I_{2}=2J^{a}P_{a}+Q_{\alpha a_{1}\dots a_{n}}C^{\alpha\beta}Q_{\beta}^{a_{1}\dots a_{n}}.\label{eq:Casimir-n}
\end{equation}
It is also worth pointing out that the super-Poincaré algebra corresponds
to the case of $n=0$, while the hyper-Poincaré algebra described
above is recovered for $n=1$.

\subsection{Hypergravity in the generic case}

The minimal coupling of General Relativity with a massless fermionic
field of spin $s=n+\frac{3}{2}$, described by a completely symmetric
$\Gamma$-traceless 1-form $\psi_{a_{1}\dots a_{n}}$, can then be
formulated in terms of a gauge field for the hyper-Poincaré algebra,
which now reads
\begin{equation}
A=e^{a}P_{a}+\omega^{a}J_{a}+\psi_{a_{1}\dots a_{n}}^{\alpha}Q_{\alpha}^{a_{1}\dots a_{n}}\,.\label{eq:Connection-n}
\end{equation}
The components of the curvature 2-form are then given by
\begin{equation}
F=R^{a}J_{a}+\tilde{T}^{a}P_{a}+D\psi_{a_{1}\dots a_{n}}^{\alpha}Q_{\alpha}^{a_{1}\dots a_{n}},\label{eq:Curvature-n}
\end{equation}
where the covariant derivative of the spin-$\left(n+\frac{3}{2}\right)$
field can be written as
\begin{equation}
D\psi^{a_{1}\dots a_{n}}=d\psi^{a_{1}\dots a_{n}}+\left(n+\frac{1}{2}\right)\omega_{b}\Gamma^{b}\psi^{a_{1}\dots a_{n}}-\omega_{b}\Gamma^{(a_{1}}\psi^{a_{2}\dots a_{n})b}\,,
\end{equation}
and 
\begin{equation}
\tilde{T}^{a}=T^{a}-\frac{1}{2}\left(n+\frac{1}{2}\right)i\bar{\psi}_{a_{1}\dots a_{n}}\Gamma^{a}\psi^{a_{1}\dots a_{n}}\,.
\end{equation}
The transformation rules of the fields under local hypersymmetry can
then be obtained from a gauge transformation of the connection \eqref{eq:Connection-n}
with a fermionic parameter given by $\lambda=\epsilon_{a_{1}\dots a_{n}}^{\alpha}Q_{\alpha}^{a_{1}\dots a_{n}}$,
so that they read
\begin{eqnarray}
\delta e^{a} & = & \left(n+\frac{1}{2}\right)i\bar{\epsilon}_{a_{1}\dots a_{n}}\Gamma^{a}\psi^{a_{1}\dots a_{n}}\,,\nonumber \\
\delta\omega^{a} & = & 0\,,\label{eq:trans-n}\\
\delta\psi^{a_{1}\dots a_{n}} & = & D\epsilon^{a_{1}\dots a_{n}}\,.\nonumber 
\end{eqnarray}
The Casimir operator \eqref{eq:Casimir-n} then implies the existence
of an (anti-) symmetric tensor of rank 2, which once contracted with
the wedge product of two curvatures, gives
\begin{eqnarray}
\left\langle F^{2}\right\rangle  & = & 2R^{a}\tilde{T}_{a}+iD\bar{\psi}_{a_{1}\dots a_{n}}D\psi^{a_{1}\dots a_{n}}\nonumber \\
 & = & d\left(2R^{a}e_{a}+i\bar{\psi}_{a_{1}\dots a_{n}}D\psi^{a_{1}\dots a_{n}}\right)\,,\label{eq:Invariant-Bilinear-1}
\end{eqnarray}
being an exact form that is manifestly invariant under the hypersymmetry
transformations \eqref{eq:trans-n}. Therefore, as in the case of
(super)gravity \cite{Achucarro:1987vz}, \cite{Witten:1988hc}, the
action can also be written as a Chern-Simons one \eqref{eq:CS-action},
which up to a boundary term reduces to
\begin{equation}
I=\frac{k}{4\pi}\int2R^{a}e_{a}+i\bar{\psi}_{a_{1}\dots a_{n}}D\psi^{a_{1}\dots a_{n}}\,,
\end{equation}
so that the field equations now read $F=0$, with $F$ given by \eqref{eq:Curvature-n}. 

Note that the standard supergravity action in \cite{Deser:1982sw},
\cite{Deser:1982sv}, \cite{Marcus:1983hb} is recovered for $n=0$;
and as it occurs in the spin-$\frac{5}{2}$ case, the generic theory
agrees with the one of Aragone and Deser only on-shell.

We would like to stress that a deeper understanding of the theory
cannot be attained unless it is endowed with a consistent set of boundary
conditions. In this sense, one of the advantages of formulating hypergravity
as a Chern-Simons theory is that the analysis of its asymptotic structure
can be directly performed in a canonical form, as in the case of negative cosmological
constant \cite{Hyper-AdS}. Indeed, in analogy with the case of three-dimensional
flat supergravity \cite{BDMT}, the mode expansion of the asymptotic
symmetry algebra of hypergravity with a spin-$\frac{5}{2}$ fermionic
field is defined through the following Poisson brackets \cite{FMT2}
\begin{eqnarray}
i\left\{ \mathcal{J}_{m},\mathcal{J}_{n}\right\}  & = & \left(m-n\right)\mathcal{J}_{m+n}\,,\nonumber \\
i\left\{ \mathcal{J}_{m},\mathcal{P}_{n}\right\}  & = & \left(m-n\right)\mathcal{P}_{m+n}+km\left(m^{2}-1\right)\delta_{m+n,0}\,,\nonumber \\
i\left\{ \mathcal{P}_{m},\mathcal{P}_{n}\right\}  & = & 0\quad,\quad i\left\{ \mathcal{P}_{m},\psi_{n}\right\} =0\,,\nonumber \\
i\left\{ \mathcal{J}_{m},\psi_{n}\right\}  & = & \left(\frac{3m}{2}-n\right)\psi_{m+n}\,,\label{eq:hyper-BMS}\\
i\left\{ \psi_{m},\psi_{n}\right\}  & = & \frac{1}{4}\left(6m^{2}-8mn+6n^{2}-9\right)\mathcal{P}_{m+n}+\frac{9}{4k}\sum_{q}\mathcal{P}_{m+n-q}\mathcal{P}_{q}\nonumber \\
 &  & +k\left(m^{2}-\frac{9}{4}\right)\left(n^{2}-\frac{1}{4}\right)\delta_{m+n,0}\,,\nonumber 
\end{eqnarray}
which describe a nonlinear hypersymmetric extension of the BMS$_{3}$
algebra \cite{ABS-BMS}, \cite{Barnich-Compere}, \cite{Barnich-Cedric}.
It can also be shown that this algebra corresponds to a subset of
a suitable contraction of two copies of the \textrm{WB}$_2$ algebra
\cite{FST-W1}, \cite{BKS-W2}, \cite{Hyper-AdS}.

When fermions fulfill antiperiodic boundary conditions, the modes
of the fermionic global charges $\psi_{m}$ are labelled by
half-integers, so that the wedge algebra of \eqref{eq:hyper-BMS}
reduces to the one of the hyper-Poincaré group. In fact, dropping
the nonlinear terms, and restricting the modes according to $|n|<\Delta$,
where $\Delta$ stands for the conformal weight of the generators,
the hyper-Poincaré algebra is manifestly recovered provided the modes
in \eqref{eq:hyper-BMS} are identified with the generators $J_{a},\, P_{a},\, Q_{\alpha a}$,
according to
\begin{gather}
\mathcal{J}_{-1}=-2J_{0}\quad,\quad\mathcal{J}_{1}=J_{1}\quad,\quad\mathcal{J}_{0}=J_{2}\,,\nonumber \\
\mathcal{P}_{-1}=-2P_{0}\quad,\quad\mathcal{P}_{1}=P_{1}\quad,\quad\mathcal{P}_{0}=P_{2}\,,\nonumber \\
\psi_{-\frac{3}{2}}=2^{\frac{5}{4}}\sqrt{3}Q_{+0}\quad,\quad \psi_{-\frac{1}{2}}=2^{\frac{3}{4}}\sqrt{3}Q_{-0}\,,\\
\psi_{\frac{1}{2}}=-2^{\frac{1}{4}}\sqrt{3}Q_{+1}\quad,\quad \psi_{\frac{3}{2}}=-2^{-\frac{1}{4}}\sqrt{3}Q_{-1}\,.\nonumber 
\end{gather}

It is also worth noting that \eqref{eq:hyper-BMS} can then be regarded
as a hypersymmetric extension of the Galilean conformal algebra in
two dimensions \cite{Bagchi}, \cite{Bagchi-Gopakumar}, which is
isomorphic to BMS$_{3}$ and turns out to be relevant in the context
of non-relativistic holography.

Another advantage of formulating hypergravity in terms of a Chern-Simons
action is that, as in case of supergravity \cite{GTW-Reloaded}, \cite{BDMT},
the theory can be readily extended to include parity odd terms in
the Lagrangian. This can be explicitly performed by a simple modification
of the invariant bilinear form, so that it acquires an additional
component given by $\left\langle J_{a},J_{b}\right\rangle =\mu\eta_{ab}$,
followed by a shift in the spin connection of the form $\omega^{a}\rightarrow\omega^{a}+\gamma e^{a}$,
so that the constants $\mu$, $\gamma$ parametrize the new allowed
couplings in the action. As a consequence, when hypergravity is extended
in this way, the hyper-BMS$_{3}$ algebra \eqref{eq:hyper-BMS} acquires
an additional nontrivial central extension along its Virasoro subgroup.

\section{Ending remarks}

The hyper-Poincaré group admits a consistent generalization to the
case of $d\geq3$ spacetime dimensions. In the case of fermionic $\Gamma$-traceless
spin-$\frac{3}{2}$ generators, the nonvanishing (anti-) commutators
of the algebra are given by
\begin{eqnarray}
\left[J_{ab},J_{cd}\right] & = & J_{ad}\eta_{bc}-J_{bd}\eta_{ac}+J_{ca}\eta_{ad}\,,\nonumber \\
\left[J_{ab},P_{c}\right] & = & P_{a}\eta_{bc}-P_{b}\eta_{ac}\,,\nonumber \\
\left[J_{ab},Q_{c}^{\alpha}\right] & = &- \frac{1}{2}(\Gamma_{ab})_{\,\,\,\,\beta}^{\alpha}Q_{c}^{\beta}+Q_{a}^{\alpha}\eta_{bc}-Q_{b}^{\alpha}\eta_{ac}\,,\\
\left[J_{ab},\bar{Q}_{\alpha c}\right] & = & \frac{1}{2}(\Gamma_{ab})_{\,\,\,\,\alpha}^{\beta}\bar{Q}_{\beta c}+\bar{Q}_{\alpha a}\eta_{bc}-\bar{Q}_{\alpha b}\eta_{ac}\,,\nonumber \\
\left\{ Q^{\alpha a},\bar{Q}_{\beta}^{b}\right\}  & = & \frac{3\left(d-2\right)}{d^{2}}i\left[\left(d+1\right)(\Gamma^{c})_{\,\,\,\,\beta}^{\alpha}P_{c}\eta^{ab}-\frac{d+2}{d-2}(\Gamma^{abc})_{\,\,\,\,\beta}^{\alpha}P_{c}-(\Gamma^{\left(a\right|})_{\,\,\,\,\beta}^{\alpha}P^{\left|b\right)}\right]\,,\nonumber 
\end{eqnarray}
where $\bar{Q}_{a}=Q_{a}^{\dagger}\Gamma^{0}$ stands
for the Dirac conjugate.

In the generic case, the spin-$\left(n+\frac{1}{2}\right)$ generators
correspond to completely symmetric $\Gamma$-traceless tensor-spinors
that fulfill $\Gamma^{a_{1}}Q_{a_{1}\dots a_{n}}=0$. In order to
avoid the intricacies related to the latter condition, as well as
with the suitable (anti-) symmetrization of the (anti-) commutation
rules of the generators, it is better to express the algebra in terms
of its Maurer-Cartan form. It is now given by
\begin{equation}
\Omega=\rho^{a}P_{a}+\frac{1}{2}\tau^{ab}J_{ab}+\bar{\chi}_{\alpha}^{a_{1}\dots a_{n}}Q_{a_{1}\dots a_{n}}^{\alpha}-\bar{Q}_{\alpha}^{a_{1}\dots a_{n}}\chi_{a_{1}\dots a_{n}}^{\alpha}\,,\label{eq:MC-form-1-1}
\end{equation}
where $\chi_{a_{1}\dots a_{n}}$ is $\Gamma$-traceless and completely
symmetric in the vector indices, so that its components fulfill %
\footnote{In the case of $d=2$ spacetime dimensions the algebra is consistent.
However, the subset spanned by translations and the fermionic generators
is an abelian ideal. %
}
\begin{align}
d\tau^{ab} & =-\tau_{\,\,\,\, c}^{a}\tau^{cb}\,,\label{eq:MC-algebra-1}\\
d\rho^{a} & =-\tau_{\,\,\,\, b}^{a}\rho^{b}+\frac{1}{2}\left(n+\frac{1}{2}\right)i\bar{\chi}_{a_{1}\dots a_{n}}\Gamma^{a}\chi^{a_{1}\dots a_{n}}\,,\nonumber \\
d\chi^{a_{1}\dots a_{n}} & =-\frac{1}{4}\tau^{ab}\Gamma_{ab}\chi^{a_{1}\dots a_{n}}-\tau_{\quad\,\,\,\, b}^{(a_{1}}\chi^{a_{2}\dots a_{n})b}\,,\nonumber \\
d\bar{\chi}^{a_{1}\dots a_{n}} & =-\frac{1}{4}\bar{\chi}^{a_{1}\dots a_{n}}\tau^{ab}\Gamma_{ab}-\tau_{\quad\,\,\,\, b}^{(a_{1}}\bar{\chi}^{a_{2}\dots a_{n})b}\,.\nonumber 
\end{align}
This algebra can be easily written in terms of Majorana spinors when
they exist, and it reduces to super-Poincaré for $n=0$.

Note that there was no need to enlarge the Lorentz group in order
to accommodate the higher spin generators, so that the additional
symmetries do not seem to interfere with the causal structure. Indeed,
as in the case of supersymmetry, the quotient of the hyper-Poincaré
group over the Lorentz subgroup now defines a hyperspace which is
an extension of Minkowski spacetime with additional $\Gamma$-traceless
tensor-spinor coordinates. However, as anticipated by Haag, \L opusza\'{n}ski
and Sohnius, the irreducible representations, which could be obtained
from suitable hyperfields, necessarily contain higher spin fields.
Nevertheless, it would be worth to explore whether the hyper-Poincaré
algebra may manifest itself through theories or models whose fundamental
fields do not transform as linear multiplets, as it would be the case
of nonlinear realizations, hyper-Poincaré-valued gauge fields, or
extended objects.

\acknowledgments We thank A. Campoleoni, M. Henneaux, G. Lucena Gómez,
A. Pérez, R. Rahman and D. Tempo for helpful discussions and enlightening
comments. O.F. and R.T. thank the International Solvay Institutes
and the ULB for warm hospitality. O.F. thanks Conicyt for financial
support. This research has been partially supported by Fondecyt grants
Nº 1130658, 1121031, 3150448. The Centro de Estudios Científicos (CECs)
is funded by the Chilean Government through the Centers of Excellence
Base Financing Program of Conicyt.


\begin{thebibliography}{10}
\bibitem{FF-SUSY}P.~Fayet and S.~Ferrara,   ``Supersymmetry,''   
Phys.\ Rept.\  {\bf 32}, 249 (1977).   

\bibitem{Sohnius-SUSY} M.~F.~Sohnius,   ``Introducing Supersymmetry,''   
Phys.\ Rept.\  {\bf 128}, 39 (1985).   

\bibitem{VN-SUGRA} P.~Van Nieuwenhuizen,   ``Supergravity,''   
Phys.\ Rept.\  {\bf 68}, 189 (1981).   

\bibitem{HLS}R.Haag, J.T.Lopuszanski and M.Sohnius, ``All Possible Generators of Supersymmetries of the s Matrix,'' 
Nucl.\ Phys.\ B 88, 257 (1975). 

\bibitem{Weinberg}  S.~Weinberg,   ``Photons and Gravitons in s Matrix Theory: Derivation of Charge Conservation and Equality of Gravitational and Inertial Mass,''   
Phys.\ Rev.\  {\bf 135}, B1049 (1964).   

\bibitem{AD-ConsistencyProb}C.~Aragone and S.~Deser,   ``Consistency Problems of Hypergravity,''   
Phys.\ Lett.\ B {\bf 86}, 161 (1979).   

\bibitem{BVHVN} F.~A.~Berends, J.~W.~van Holten, P.~van Nieuwenhuizen and B.~de Wit,   ``On Spin 5/2 Gauge Fields,''   Phys.\ Lett.\ B {\bf 83}, 188 (1979)   [Phys.\ Lett.\  {\bf 84B}, 529 (1979)].   

\bibitem{AD2} C.~Aragone and S.~Deser,   ``Higher Spin Vierbein Gauge Fermions and Hypergravities,''   Nucl.\ Phys.\ B {\bf 170}, 329 (1980).   

\bibitem{Weinberg-Witten}  S.~Weinberg and E.~Witten,   ``Limits on Massless Particles,''   
Phys.\ Lett.\ B {\bf 96}, 59 (1980).   

\bibitem{Porrati} M.~Porrati,   ``Universal Limits on Massless High-Spin Particles,''   
Phys.\ Rev.\ D {\bf 78}, 065016 (2008)   [arXiv:0804.4672 [hep-th]].   

\bibitem{Bekaert-Boulanger-Sundell}X.~Bekaert, N.~Boulanger and P.~Sundell,   ``How higher-spin gravity surpasses the spin two barrier: no-go theorems versus yes-go examples,''   
Rev.\ Mod.\ Phys.\  {\bf 84}, 987 (2012)   [arXiv:1007.0435 [hep-th]].   

\bibitem{Blencowe}M.P.Blencowe, ``A Consistent Interacting Massless Higher Spin Field Theory in $D$ = (2+1),'' 
Class.\ Quant.\ Grav.\ \textbf{6},443 (1989). 

\bibitem{BBS}E.Bergshoeff, M.P.Blencowe and K.S.Stelle, ``Area Preserving Diffeomorphisms and Higher Spin Algebra,'' 
Commun.\ Math.\ Phys.\ \textbf{128},213 (1990). 

\bibitem{Vasiliev-HS}  M.~A.~Vasiliev,   ``Higher spin gauge theories in four-dimensions, three-dimensions, and two-dimensions,''   
Int.\ J.\ Mod.\ Phys.\ D {\bf 5}, 763 (1996)   [hep-th/9611024].   

\bibitem{Henneaux-Rey}  M.~Henneaux and S.~J.~Rey,   ``Nonlinear $W_{infinity}$ as Asymptotic Symmetry of Three-Dimensional Higher Spin Anti-de Sitter Gravity,''   
JHEP {\bf 1012}, 007 (2010)   [arXiv:1008.4579 [hep-th]].   

\bibitem{CFPT-HS3D}  A.~Campoleoni, S.~Fredenhagen, S.~Pfenninger and S.~Theisen,   ``Asymptotic symmetries of three-dimensional gravity coupled to higher-spin fields,''  
JHEP {\bf 1011}, 007 (2010)   [arXiv:1008.4744 [hep-th]].   

\bibitem{HPTT-CPHS3D}  M.~Henneaux, A.~Perez, D.~Tempo and R.~Troncoso,   ``Chemical potentials in three-dimensional higher spin anti-de Sitter gravity,''   
JHEP {\bf 1312}, 048 (2013)   [arXiv:1309.4362 [hep-th]].   

\bibitem{BHPTT-GBH3D}  C.~Bunster, M.~Henneaux, A.~Perez, D.~Tempo and R.~Troncoso,   ``Generalized Black Holes in Three-dimensional Spacetime,''  
JHEP {\bf 1405}, 031 (2014)   [arXiv:1404.3305 [hep-th]].   

\bibitem{ABFGR-HSF3D}H.~Afshar, A.~Bagchi, R.~Fareghbal, D.~Grumiller and J.~Rosseel,   ``Spin-3 Gravity in Three-Dimensional Flat Space,''   
Phys.\ Rev.\ Lett.\  {\bf 111}, 121603 (2013)   [arXiv:1307.4768 [hep-th]].   

\bibitem{GMPT-HSF3D}  H.~A.~Gonzalez, J.~Matulich, M.~Pino and R.~Troncoso,   ``Asymptotically flat spacetimes in three-dimensional higher spin gravity,''   
JHEP {\bf 1309} (2013) 016   [arXiv:1307.5651 [hep-th]].   

\bibitem{GGRR-CPHSF3D} M.~Gary, D.~Grumiller, M.~Riegler and J.~Rosseel,   ``Flat space (higher spin) gravity with chemical potentials,''   
JHEP {\bf 1501}, 152 (2015)   [arXiv:1411.3728 [hep-th]].   

\bibitem{MPTT-CPHSF3D}  J.~Matulich, A.~Perez, D.~Tempo and R.~Troncoso,   ``Higher spin extension of cosmological spacetimes in 3D: asymptotically flat behaviour with chemical potentials and thermodynamics,''   
JHEP {\bf 1505}, 025 (2015)   [arXiv:1412.1464 [hep-th]].   

\bibitem{Hietarinta}  J.~Hietarinta,   ``Supersymmetry Generators of Arbitrary Spin,''   
Phys.\ Rev.\ D {\bf 13}, 838 (1976).   

\bibitem{AD-3D-HYGRA} C.~Aragone and S.~Deser,   ``Hypersymmetry in $D=3$ of Coupled Gravity Massless Spin 5/2 System,''
  Class.\ Quant.\ Grav.\  {\bf 1}, L9 (1984).   

\bibitem{BTZ-HDCS} M.~Banados, R.~Troncoso and J.~Zanelli,   ``Higher dimensional Chern-Simons supergravity,''   
Phys.\ Rev.\ D {\bf 54}, 2605 (1996)   [gr-qc/9601003].   

\bibitem{CLW}  B.~Chen, J.~Long and Y.~N.~Wang,   ``Conical Defects, Black Holes and Higher Spin (Super-)Symmetry,''   
JHEP {\bf 1306}, 025 (2013)   [arXiv:1303.0109 [hep-th]].   

\bibitem{ZInoviev} Y.~M.~Zinoviev, ``Hypergravity in AdS$_3$,''   
Phys.\ Lett.\ B {\bf 739}, 106 (2014)  [hep-th/1408.2912].

\bibitem{Hyper-AdS}M.~Henneaux, A.~Perez, D.~Tempo and R.~Troncoso,
  ``Hypersymmetry bounds and three-dimensional higher-spin black holes,''
  JHEP {\bf 1508}, 021 (2015)
  [arXiv:1506.01847 [hep-th]].

\bibitem{Achucarro:1987vz} A.~Achucarro and P.~K.~Townsend,   ``A Chern-Simons Action for Three-Dimensional anti-De Sitter Supergravity Theories,''
Phys.\ Lett.\ B {\bf 180}, 89 (1986).   

\bibitem{Witten:1988hc}  E.~Witten,   ``(2+1)-Dimensional Gravity as an Exactly Soluble System,'' 
 Nucl.\ Phys.\ B {\bf 311}, 46 (1988).   

\bibitem{Deser:1982sw} S.~Deser and J.~H.~Kay,   ``Topologically Massive Supergravity,''   
Phys.\ Lett.\ B {\bf 120}, 97 (1983).   

\bibitem{Deser:1982sv}  S.~Deser,   ``Cosmological Topological Supergravity,''   
In *Christensen, S.m. ( Ed.): Quantum Theory Of Gravity*, 374-381   

\bibitem{Marcus:1983hb}N.~Marcus and J.~H.~Schwarz,   ``Three-Dimensional Supergravity Theories,''
  Nucl.\ Phys.\ B {\bf 228}, 145 (1983).   

\bibitem{BDMT}G.~Barnich, L.~Donnay, J.~Matulich and R.~Troncoso,   ``Asymptotic symmetries and dynamics of three-dimensional flat supergravity,''   
JHEP {\bf 1408}, 071 (2014)   [arXiv:1407.4275 [hep-th]].   


\bibitem{FMT2}  O.~Fuentealba, J.~Matulich and R.~Troncoso,
  ``Asymptotically flat structure of hypergravity in three spacetime dimensions,''
  JHEP {\bf 1510}, 009 (2015)
  [arXiv:1508.04663 [hep-th]].

\bibitem{ABS-BMS}A.~Ashtekar, J.~Bicak and B.~G.~Schmidt,   ``Asymptotic structure of symmetry reduced general relativity,''   
Phys.\ Rev.\ D {\bf 55}, 669 (1997)   [gr-qc/9608042].   

\bibitem{Barnich-Compere} G.~Barnich and G.~Compere,   ``Classical central extension for asymptotic symmetries at null infinity in three spacetime dimensions,''   
Class.\ Quant.\ Grav.\  {\bf 24}, F15 (2007)   [gr-qc/0610130].   

\bibitem{Barnich-Cedric} G.~Barnich and C.~Troessaert,   ``Aspects of the BMS/CFT correspondence,''   
JHEP {\bf 1005}, 062 (2010)   [arXiv:1001.1541 [hep-th]].   

\bibitem{FST-W1}J.~M.~Figueroa-O'Farrill, S.~Schrans and K.~Thielemans,   ``On the Casimir algebra of B(2),''   
Phys.\ Lett.\ B {\bf 263}, 378 (1991).   

\bibitem{BKS-W2}S.~Bellucci, S.~Krivonos and A.~S.~Sorin,   ``Linearizing W(2,4) and WB(2) algebras,''   
Phys.\ Lett.\ B {\bf 347}, 260 (1995)   [hep-th/9411168].   

\bibitem{Bagchi} A.~Bagchi,   ``Correspondence between Asymptotically Flat Spacetimes and Nonrelativistic Conformal Field Theories,''   
Phys.\ Rev.\ Lett.\  {\bf 105}, 171601 (2010).   

\bibitem{Bagchi-Gopakumar}A.~Bagchi and R.~Gopakumar,   ``Galilean Conformal Algebras and AdS/CFT,''   
JHEP {\bf 0907}, 037 (2009)   [arXiv:0902.1385 [hep-th]].   


\bibitem{GTW-Reloaded}  A.~Giacomini, R.~Troncoso and S.~Willison,   ``Three-dimensional supergravity reloaded,''   
Class.\ Quant.\ Grav.\  {\bf 24}, 2845 (2007)   [hep-th/0610077].   
\end{thebibliography}
\end{document}